\documentclass[aps,prd,twocolumn,10pt,tightenlines,superscriptaddress,
nofootinbib,showpacs]{revtex4-1}
\usepackage{amssymb,latexsym}
\usepackage{amsmath,amsbsy,bbm}
\usepackage{graphicx,comment}
\usepackage{epsfig,bm}
\usepackage{slashed}
\usepackage{lipsum,color}
\newcommand\beq{\begin{eqnarray}}
\newcommand\eeq{\end{eqnarray}}

\newcommand{\cL}{\mathcal{L}}

\newcommand{\cO}{\mathcal{O}}

\def\sumint{\hat{\sum} \hskip-1.35em \int_{ \hskip-0.25em \underset{\bm{q}}{\phantom{a}} } \, \, \,}

\def\Dslash{D\hskip-0.65em /}
\def\ol#1{{\overline{#1}}}

\begin{document}


\title{Finite Volume Corrections to the Electromagnetic Mass of Composite Particles}

\author{Jong-Wan~Lee}
\email[]{jwlee2@ccny.cuny.edu}
\altaffiliation[\\Present address: ]
{Department of Physics, College of Science, Swansea University, Swansea SA2 8PP, United Kingdom}
\affiliation{
Department of Physics,
        The City College of New York,
        New York, NY 10031, USA}
\author{Brian~C.~Tiburzi}
\email[]{btiburzi@ccny.cuny.edu}
\affiliation{
Department of Physics,
        The City College of New York,
        New York, NY 10031, USA}
\affiliation{
Graduate School and University Center,
        The City University of New York,
        New York, NY 10016, USA}
\affiliation{
RIKEN BNL Research Center,
        Brookhaven National Laboratory,
        Upton, NY 11973, USA}
\date{\today}


\begin{abstract}

The long-range electromagnetic interaction presents a challenge for numerical computations in
QCD + QED. 
In addition to power-law finite volume effects, 
the standard lattice gauge theory approach introduces non-locality through removal of photon zero-momentum modes.
The resulting finite volume effects must be quantitatively understood; 
and, 
to this end,  
non-relativistic effective field theories are an efficient tool,
especially in the case of composite particles. 
Recently an oddity related to non-locality of the standard lattice approach was uncovered by the 
Budapest-Marseille-Wuppertal 
collaboration. 
Explicit contributions from antiparticles appear to be required so that finite volume QED results for a point-like fermion can be reproduced in the effective field theory description.  
We 
provide transparency for  
this argument by considering point-like scalars and spinors in finite volume QED using the method of regions. 
For the more germane case of composite particles, 
we determine that antiparticle modes contribute to the finite-volume electromagnetic mass of composite spinors
through terms proportional to the squares of time-like 
form factors evaluated at threshold. 
We extend existing finite volume calculations to one order higher, 
which is particularly relevant for the electromagnetic mass of light nuclei. 
Additionally, 
we verify that the analogous finite volume contributions to the nucleon mass in chiral perturbation theory vanish
in accordance with locality.

\end{abstract}
\maketitle

\section{Introduction}


Lattice QCD computations have progressed to the level of providing a variety of quantitative input for strong interaction phenomenology. 
Refinements to techniques and advances in both algorithms and hardware continue to be made, 
and this has rendered high-precision lattice determination of some observables in need of QED corrections. 
Such corrections, 
moreover, 
are important in addressing the light quark mass difference from first principles, 
computing the nucleon mass splitting, 
and accounting for charged hadron scattering at low energies. 
Each of these is a milestone in the further development of lattice QCD. 
Since the original investigation of QED effects on the spectrum of hadrons%
~\cite{Duncan:1996xy},
there have been considerable improvements in recent lattice calculations%
~\cite{Blum:2007cy,Blum:2010ym,Aoki:2012st,Ishikawa:2012ix,Borsanyi:2013lga,deDivitiis:2013xla,Horsley:2015eaa}, 
including an impressive computation of the hadron spectrum that dynamically includes all sources of strong and electromagnetic isospin breaking%
~\cite{Borsanyi:2014jba}.

In tandem with these lattice QCD + QED calculations, 
there has also been a focus on addressing theoretical aspects of the method
and applications beyond spectroscopy, 
for recent progress in these directions, 
see, 
for example,%
~\cite{Beane:2014qha,Davoudi:2014qua,Fodor:2015pna,Carrasco:2015xwa}. 
In this work, 
we focus on QED corrections to hadron masses at leading order in the fine-structure constant, 
$\alpha$.  
Specifically we investigate the pernicious finite volume effects that must be quantitatively 
understood in order to extract physical results. 
These effects arise from altering the long-range behavior of the electromagnetic interaction 
on a torus, 
and our analysis specializes to the non-local formulation of finite volume electrodynamics, 
in which photon zero-momentum modes are removed.%
\footnote{
A local formulation of finite volume QCD + QED using 
$C^*$
boundary conditions%
~\cite{Wiese:1991ku}
has quite recently been proposed%
~\cite{Lucini:2015hfa}, 
and shows considerable promise at ameliorating the size of finite volume corrections.
} 
An ideal framework for these computations is that of non-relativistic QED
(NRQED)%
~\cite{Caswell:1985ui,Kinoshita:1995mt,Manohar:1997qy}, 
which has recently been applied to composite non-relativistic particles, 
such as the nucleon%
~\cite{Hill:2012rh}.

Within an NRQED framework, 
finite volume corrections to the electromagnetic masses of hadrons have been derived%
~\cite{Davoudi:2014qua}.
In the case of a point-like spin-half particle, 
results derived in QED disagree with those from NRQED, 
and this is supported by numerical data%
~\cite{Borsanyi:2014jba}.
Quite recently%
~\cite{Fodor:2015pna}, 
it has been argued that the standard NRQED description is missing explicit contributions from antiparticles in finite volume on account of the non-locality.
In the present note, 
we provide transparency for this argument by considering the electromagnetic mass of point-like scalar and point-like spinor particles in QED
utilizing the method of regions;
in the present context,
see%
~\cite{Smirnov:1990rz,Beneke:1997zp}. 
Additionally we extend these considerations to composite scalar and spinor particles, 
and find there are no modifications to the finite volume effects from antiparticles in the former case, 
while there are
$\cO(\alpha L^{-3})$
effects in the latter 
that 
depend on squares of time-like form factors evaluated at the pair-production threshold. 
Existing finite volume calculations are trivially extended to one order higher.
The absence of such corrections is useful to know, 
for example, in assessing the effect of finite volume on the electromagnetic mass of light nuclei. 
We comment briefly on the nucleon-antinucleon contact operators in chiral perturbation theory, 
however, 
their finite volume contributions to the nucleon mass are shown to vanish due to locality of finite volume QCD.

The organization of this note is as follows. 
First in Sec.~\ref{sec:demo}, 
we consider the case of point-like particles in QED. 
For scalars and spinors, 
we determine the finite-volume electromagnetic mass to 
leading order in 
$\alpha$,
including both power-law and exponential contributions. 
These results are contrasted with those obtained by the standard non-relativistic framework, 
in which the power-law contributions should be reproduced. 
Disagreements are linked to antiparticle contributions, 
as we exhibit by employing the 
(asymptotic)
method of regions.  
Including particle-antiparticle contact interactions in the non-relativistic effective field theory settles the 
disagreement. 
These observations are then utilized to extend to the more germane case of 
composite scalars and composite spinors in 
Sec.~\ref{sec:comp}. 
Finite volume effects for these composite particles are determined to 
$\cO(\alpha L^{-5})$.
In the scalar case, 
there are no antiparticle contributions at finite volume, 
whereas for the spinor case, 
antiparticle modes show up exclusively at 
$\cO(\alpha L^{-3})$, 
and are completely dependent on the non-locality introduced by subtracting the photon zero mode. 
By extending the point-like spinor case considered in%
~\cite{Fodor:2015pna}
to that of a composite spinor, 
we find that the required volume effect from anitparticle modes depends on the square of the time-like Sachs magnetic form factor evaluated at threshold.  
We address related concerns in the finite volume computation of the nucleon mass using chiral perturbation theory in 
Sec.~\ref{sec:chiral}.
Antiparticle modes are argued not to make a contribution to the volume dependence of the nucleon mass due to locality of finite volume QCD. 
Our findings are summarized in 
Sec.~\ref{sec:summary}, 
details about the one-loop computations for point-like particles are contained in 
Appendix~\ref{A:A},
and shape coefficients used throughout are evaluated in 
Appendix~\ref{A:B}.

\section{Point-Like Particles}
\label{sec:demo}

To establish the need for antiparticle contributions,
we evaluate the finite volume corrections to the electromagnetic mass of charged, point-like particles.
We consider both scalars and spinors in QED to leading order in the fine-structure constant.  
These results are contrasted with the non-relativistic approach, 
and discrepancies in the power-law corrections are transparently linked to antiparticle contributions by employing the method of regions.

\subsection{One-Loop Computations}

In all of our computations, 
we consider the theory of QED defined on a three-dimensional spatial torus of length 
$L$ 
in each direction, 
with the time direction kept infinite.  
To avoid singularities associated with the long-range nature of the electromagnetic interaction, 
we introduce non-locality by removing the three-momentum zero modes of the photon. 
The effect of this non-locality surfaces in finite volume corrections. 
We work in the particle's rest frame, 
$p_\mu = (M, \bm{0})$. 
Results in moving frames are not identical because the finite volume breaks Lorentz invariance.

\subsubsection{Scalar QED}

\begin{figure}
\epsfig{file=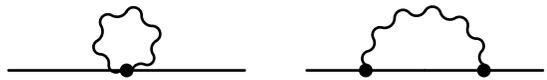,width=0.4\textwidth}
\caption{
Electromagnetic mass corrections at $\mathcal{O}(\alpha)$.
Circles denote the relativistic charge and charge-squared couplings of the photon. 
The latter coupling is only present in the case of scalar electrodynamics.}
\label{f:sun}
\end{figure}

The relativistic scalar QED Lagrangian density has the form 
\begin{equation}
\cL
=
D^\mu \Phi^\dagger D_\mu \Phi - M^2 \Phi^\dagger \Phi
,\end{equation}
where 
$D_\mu \Phi = (\partial_\mu + i Q e A_\mu) \Phi$,
along with
$D_\mu \Phi^\dagger =  (\partial_\mu - i Q e A_\mu) \Phi^\dagger$,
and 
$e$
is the magnitude of the electron's charge. 
To leading order in the fine-structure constant 
$\alpha$
and to all orders in the particle's Compton wavelength
$M^{-1}$,
the long-range part of the electromagnetic mass is determined by the sunset and tadpole diagrams shown in 
Fig.~\ref{f:sun}.  
Of course there are also short-distance contributions required which renormalize the particle's mass.

%
%
%
%

Evaluation of the Feynman diagrams is straightforward. 
In the rest frame and with the subtraction of the photon's three-momentum zero modes, 
we arrive at the following intermediate expression for the finite volume electromagnetic mass shift
\begin{eqnarray}
\Delta M_\Phi
&=& 
- i Q^2 e^2
\int \frac{d q_0}{2 \pi}
\sumint
\frac{1}{q^2 + i \epsilon}
\notag \\
&& \phantom{space}
\times
\Bigg[
\frac{2 M + q_0}{2 M q_0 + q^2 + i \epsilon}
- 
\frac{3}{2M}
\Bigg]
\label{eq:demo}
.\end{eqnarray}
Throughout, 
finite volume corrections to a quantity 
$\mathcal{M}$
are defined by 
$\Delta \mathcal{M} = \mathcal{M}(L) -  \mathcal{M}(\infty)$, 
and for brevity we employ the notation
\begin{equation}
\sumint f(\bm{q})
\equiv
\frac{1}{L^3} \sum_{\bm{n} \neq \bm{0}}
f \left( \frac{2 \pi \bm{n}}{L} \right)
- 
\int \frac{d\bm{q}}{(2 \pi)^3}
f(\bm{q})
,\end{equation}
where the hat denotes exclusion of the three-momentum zero mode.

The finite volume correction shown in 
Eq.~\eqref{eq:demo} 
evaluates to
\begin{equation}
\frac{\Delta M_\Phi}{ M}
=
\frac{\Delta M^{(1,2)}}{M}
-  
\frac{6 \pi \, Q^2 \alpha}{(2 \pi ML)^{3/2}} e^{- ML}
+ 
\cdots
\label{eq:scalarL}
,\end{equation}
where, 
for illustrative purposes, 
we have retained the leading exponential correction, 
see Appendix~\ref{A:A} for the full expression.  
The 
\emph{universal} 
power-law finite volume correction 
$\Delta M^{(1,2)}$
appearing above
is the same for all particles%
~\cite{Borsanyi:2014jba,Davoudi:2014qua},  
includes terms at 
$\cO(\alpha L^{-1})$
and 
$\cO(\alpha L^{-2})$, 
and has the explicit form
\begin{equation}
\frac{\Delta M^{(1,2)}}{M}
=
\frac{Q^2 \alpha}{2 \pi M L}
c_2 
+ 
\frac{Q^2 \alpha}{(M L)^2}
c_1
\label{eq:M12}
.\end{equation}
The parameters 
$c_j$
above are pure numbers that reflect the modes allowed in the spatial geometry; 
and, 
for a spatial torus, 
they are defined by
\begin{equation}
c_j
=
\sum_{\bm{n} \neq \bm{0}}
\frac{1}{\, |\bm{n}|^j}
- 
\int \frac{d \bm{n}}{\, |\bm{n}|^j}
\label{eq:cjs}
,\end{equation}
see 
Appendix~\ref{A:B}
for their evaluation.

In considering non-relativistic effective field theories below, 
only the power-law finite volume dependence can be determined. 
To this end, 
it is helpful to cast the finite volume correction, 
Eq.~\eqref{eq:scalarL}, 
in the form
\begin{equation}
\frac{
\Delta M_\Phi}
{M}
=
Q^2 \alpha
\left[
\frac{c_2}{2 \pi} 
\xi 
+ 
c_1 \,
\xi^2 
-
\frac{3}{\sqrt{2 \pi}}
\xi^{\frac{3}{2}} 
\exp \left(- \frac{1}{\xi} \right)
+ 
\cdots
\right]
\label{eq:xiexp}
,\end{equation}
where the small dimensionless parameter 
$\xi$
is defined by
$\xi = (M L)^{-1}$. 
Higher-order terms neglected above all depend exponentially on the volume. 
Considering an expansion in 
$\xi$, 
the power-law corrections represent perturbative contributions that can be obtained in the effective theory, 
while exponential corrections are non-perturbative and are thus inaccessible in the non-relativistic effective theory.

\subsubsection{QED}

The interactions of a point-like spin-$\frac{1}{2}$ particle 
$\Psi$
in QED are described by the Lagrangian density
\begin{equation}
\cL 
= 
\ol \Psi \left( i \Dslash - M \right) \Psi
,\end{equation}
where 
$D_\mu \Psi = (\partial_\mu + i Q e A_\mu) \Psi$.
To leading order in the fine-structure constant 
$\alpha$
and to all orders in the particle's Compton wavelength
$M^{-1}$,
the long-range part of the electromagnetic mass is determined by the sunset diagram shown in 
Fig.~\ref{f:sun}.  
Short-distance contributions that renormalize the particle's mass are also required but not depicted.

As in the scalar case, 
evaluation of the Feynman diagram in the particle's rest frame is straightforward. 
We provide an intermediate stage of the calculation, 
at which point we have the finite volume mass shift
\begin{eqnarray}
\Delta M_\Psi
&=&
- 2 i Q^2 e^2
\int \frac{d q_0}{2 \pi}
\sumint
\frac{1}{q^2 + i \epsilon}
\, 
\frac{M - q_0}{2 M q_0 + q^2 + i \epsilon}
.\notag \\
\label{eq:demo2}
\end{eqnarray}
Carrying out the computation of the finite volume effect, 
we arrive at
\begin{equation}
\frac{
\Delta M_\Psi}{M}
=
\frac{\Delta M^{(1,2)} }{M}
+
\frac{3 \pi \, Q^2 \alpha}{(ML)^3}
-  
\frac{24 \pi \, Q^2 \alpha}{(2 \pi ML)^{3/2}} e^{- ML}
+ 
\cdots
\label{eq:pointspin}
,\end{equation}
where the universal finite volume correction is given in Eq.~\eqref{eq:M12}, 
and only the leading exponential has been shown, 
see Appendix~\ref{A:A}. 
Unlike the scalar case, 
there is a non-vanishing correction beyond the universal power-law terms 
because of the subtraction of photon's zero-momentum mode. 
The power-law corrections are identical to those obtained in%
~\cite{Borsanyi:2014jba}.

\subsection{Non-Relativistic Particles}

Now we wish to address whether the finite volume behavior for scalar and spinor particles can similarly be 
obtained using non-relativistic effective theory. 
To accomplish this, 
we shall transparently derive the non-relativistic results using the method of regions%
~\cite{Smirnov:1990rz,Beneke:1997zp}. 
The advantage of this method is a one-to-one correspondence between regions of integration and the effective degrees of freedom. 
Of course, 
the disadvantage is that this asymptotic method can only be applied in perturbation theory, 
and so will not produce exponential corrections. 
Furthermore, 
the method does not straightforwardly extend to composite particles.

We start from the relativistic expressions for the mass shift given in 
Eqs.~\eqref{eq:demo} and \eqref{eq:demo2}, 
and inspect the pole structure of the integrand.
In terms of pole structure, 
both expressions are essentially the same. 
The photon propagator has poles exclusively in the small-energy regime,
$q_0 \sim | \bm{q} | \ll M$. 
On the other hand, 
the propagator of the matter field has a pole in this small-energy regime, 
as well as a pole in the regime 
$q_0 \sim - 2 M$, 
$|\bm{q}| \ll M$, 
which corresponds to the contribution from an intermediate-state antiparticle.
To exhibit this, 
we write the propagator of the matter field as
\begin{equation}
\frac{1}{2 M q_0 + q^2 + i \epsilon}
=
\frac{1}{(q_0 - E_+ + i \epsilon)(q_0 - E_- - i \epsilon)}
,\end{equation}
with 
\begin{eqnarray}
E_+ 
&=&
- M + \sqrt{M^2 + \bm{q}^2} = \frac{\bm{q}^2}{2 M} + \cdots,
\notag \\
E_-
&=&
- M - \sqrt{M^2 + \bm{q}^2} = - 2 M - \frac{\bm{q}^2}{2M} + \cdots
,\end{eqnarray}
where 
$E_\pm$
correspond to the particle and antiparticle states, respectively.

In the small-energy region of integration, 
we can approximate the antiparticle pole by analytic terms because
$q_0 \ll |E_-|$. 
Inserting the expansion
\begin{equation}
\frac{1}{q_0 - E_- - i \epsilon}
=
- \frac{1}{E_-}
\sum_{j=0}^\infty
\left( \frac{q_0}{E_-} \right)^j
\label{eq:local}
,\end{equation}
and integrating term-by-term clearly leads to an asymptotic expansion, 
but produces the non-relativistic effective field theory answer. 
As the poles from the photon and particle state are retained, 
these are the corresponding effective degrees of freedom. 
For the scalar case, 
this non-relativistic expansion gives rise to the finite volume electromagnetic mass shift
\begin{equation}
\Delta M_\Phi^{\text{NR}}
=
\Delta M^{(1,2)}
,\end{equation}
where all higher-order power-law corrections, 
which have the form
$(M L)^{- 2 j -3}$
for 
$j = 1, 2, \cdots$,
are absent due to vanishing of the corresponding shape coefficients
$c_{- 2 j}$, 
see Appendix~\ref{A:B}.
In this way, 
we recover the finite volume effect 
that is perturbative in the parameter
$\xi$, 
see Eq.~\eqref{eq:xiexp}.

For the spinor case, 
the non-relativistic expansion gives us the finite-volume mass shift 
\begin{equation}
\frac{
\Delta M_\Psi^{\text{NR}}}
{M}
=
\frac{\Delta M^{(1,2)}}{M}
-
\frac{3 \pi \, Q^2 \alpha}{2 (M L)^3}
\, c_{0} 
,\end{equation}
in which the 
$\cO(\alpha L^{-3})$
contribution is a factor of two smaller than the correct result appearing in
Eq.~\eqref{eq:pointspin}. 
This discrepancy has been discussed in%
~\cite{Borsanyi:2014jba,Fodor:2015pna}.

Given the discrepancy,
we see transparently that the finite volume effect from antiparticles cannot be represented by the series of local terms
obtained in Eq.~\eqref{eq:local}. 
Further degrees of freedom are hence required. 
Notice that this analytic antiparticle approximation can be found by ignoring issues of convergence, 
and directly series expanding the integrands in 
Eqs.~\eqref{eq:demo} and \eqref{eq:demo2}
as a function of 
$M^{-1}$ 
before integration. 

\subsection{Non-Relativistic Antiparticles}

The discrepancy between relativistic and non-relativistic results 
can be resolved by including contributions from antiparticles. 
We first show that this is the case using the method of regions. 
Then we turn to accounting for antiparticle contributions in the effective theory.

\begin{figure}
\epsfig{file=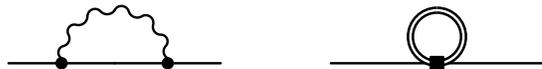,width=0.4\textwidth}
\caption{
The fully relativistic sunset diagram reduces to the antiparticle tadpole topology of the effective theory in the limit of large photon virtuality.
The antiparticle tadpoles give the contributions in 
Eqs.~\eqref{eq:antiscalar} and \eqref{eq:antispinor}
that are required for the effective field theory to reproduce the QED results.}
\label{f:anntad}
\end{figure}

Returning to Eqs.~\eqref{eq:demo} and \eqref{eq:demo2},  
we investigate the region of integration where the antiparticle contribution is dominant. 
In this region, 
we have 
$q_0 \sim - 2 M$ 
and 
$|\bm{q} | \ll M$, 
for which the particle contribution can be approximated by a tower of analytic terms. 
The photon propagator, 
moreover, 
can similarly be approximated by analytic terms in this region. 
Performing a series expansion about 
$p_0 \equiv q_0 + 2M \approx 0$,
we have the leading-order replacements
\begin{eqnarray}
(q_0 - E_- - i \epsilon)^{-1} 
&=& 
\phantom{-}
(p_0 - E_- - 2M -  i \epsilon)^{-1}
,\notag \\
(q_0 - E_+ + i \epsilon)^{-1} 
&=&
-(2M + E_+)^{-1}
+ 
\cO(p_0)
,\notag \\
(q^2 + i \epsilon)^{-1}
&=& 
\phantom{-}
(4M^2 - \bm{q}^2)^{-1}
+ 
\cO(p_0)
\label{eq:poles}
.\end{eqnarray}
Notice that 
$E_- + 2M$
vanishes for large 
$M$. 
The physical interpretation of these approximations is that only the antiparticle part of the matter field propagates in the loop.
The particle part has been replaced by analytic terms. 
This is additionally the case for the photon propagator, 
as the photon necessarily has large virtuality, 
$q^2 \geq (2 M)^2$. 
In the effective field theory description, 
we thus have the external states coupled to virtual antiparticle modes via local operators, 
see Fig.~\ref{f:anntad}.

%
%
%
%

Inserting the above approximations into the scalar electromagnetic mass shift, 
Eq.~\eqref{eq:demo}, 
and
subsequently expanding the integrand in 
$\bm{q}^2 / M^2$, 
we arrive at the leading contribution from the antiparticle pole
\begin{equation}
\Delta M_{\Phi}^{\text{anti}}
=
i \frac{Q^2 e^2}{8 M^2}
\int \frac{dp_0}{2\pi} 
\,
\sumint
\frac{p_0}{p_0 + \frac{\bm{q}^2}{2M} - i \epsilon}
\label{eq:antiscalar}
.\end{equation}
Evaluation of the integral does not produce a finite volume correction at 
$\cO(\alpha L^{-3})$, 
and explains why non-locality has no effect on point-like scalars.

For the spin-$\frac{1}{2}$ case, 
the leading antiparticle contribution takes the form
\begin{equation}
\Delta M_{\Psi}^{\text{anti}}
=
\frac{3 i Q^2 e^2}{4 M^2}
\int \frac{dp_0}{2\pi} 
\,
\sumint
\frac{1}{p_0 - i \epsilon}
\label{eq:antispinor}
,\end{equation}
on account of inserting the approximations in Eq.~\eqref{eq:poles}
into the expression for the mass shift, 
Eq.~\eqref{eq:demo2}. 
In turn, 
the missing contribution to the spinor's finite volume electromagnetic mass shift emerges
\begin{equation}
\frac{\Delta M_\Psi^{\text{anti}}}{M}
=
- 
\frac{3 \pi \, Q^2 \alpha}{2 (M L)^3}
\, c_0
\label{eq:missingspin}
,\end{equation}
which is such that
$\Delta M_\Psi^{\text{NR}} + \Delta M_\Psi^{\text{anti}} = \Delta M_\Psi$.

%
%
\begin{figure}
\epsfig{file=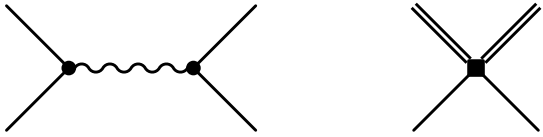,width=0.4\textwidth}
\caption{
Particle-antiparticle annihilation diagrams in the full and effective theory. 
The necessarily large photon virtuality reduces to a contact interaction in the effective theory.
}
\label{f:contact}
\end{figure}
%
%

To capture such missing contributions in an effective field theory framework, 
we thus must retain antiparticle degrees of freedom. 
We begin by considering the scalar case, 
for which we must demonstrate that such contributions are absent. 
The Lagrangian density including the non-relativistic scalar field 
$\phi$, 
and
the non-relativistic anti-scalar field 
$\chi$
schematically reads
\begin{equation}
\cL
= 
\cL_{pt.} (\phi, e)
+ 
\cL_{pt.} (\chi, -e)
+ 
\cL_c (\phi, \chi)
\label{eq:fullpt}
,\end{equation}
where 
$\cL_c(\phi, \chi)$
contains contact interactions
and is our present focus. 
The non-relativistic fields interact through the exchange of photons
given the couplings for point-like particles.  
These interactions reproduce the photon-exchange interaction of the relativistic theory order-by-order in 
$M^{-1}$, 
with one exception:
the annihilation channel.  
In this channel, 
the intermediate-state photon has large virtuality which excludes this type of interaction from the effective theory. 
For large virtuality, 
however, 
the photon propagator contracts to a point, and the annihilation process thus appears in the effective theory in the form of contact interactions between particles and antiparticles%
~\cite{Bodwin:1994jh}.  
These appear in 
$\cL_c (\phi, \chi)$. 
Matching the relativistic annihilation diagram to a contact interaction, 
see Fig.~\ref{f:contact}, 
we find
\begin{eqnarray}
\cL_c (\phi, \chi)
=
- \frac{Q^2 e^2}{16 M^4}
\left(
\phi^\dagger 
i\overset{\leftrightarrow}{\bm{D}}  \,
\chi
\right)
\cdot
\left(
\chi^\dagger  
i \overset{\leftrightarrow}{\bm{D}} \,
\phi
\right)
\label{eq:scontact}
,\end{eqnarray}
as the leading contribution in the limit of small momenta. 
In the above expression, 
we define the Hermitian derivative
$i \overset{\leftrightarrow}{D}_\mu$
by its action
$A^\dagger i \overset{\leftrightarrow}{D}_\mu B \equiv A^\dagger [ i D_\mu B ] - [i D_\mu A^\dagger] B$.
Notice from 
Eq.~\eqref{eq:poles}, 
that the limit of small momenta corresponds to expanding the photon's virtuality above the threshold for pair production.

Operators of the contact Lagrangian produce contributions to the particle's self energy through tadpole diagrams. 
The antiparticle tadpoles arise from self contraction of the antiparticle fields in the contact Lagrangian. 
These are the contributions considered in%
~\cite{Fodor:2015pna}
for a point-like spin-half particle. 
Physically the tadpole diagram takes into account the contribution to the relativistic sunset diagram in the limit of large photon virtuality, 
see Fig.~\ref{f:anntad}. 
Computation of the antiparticle tadpole diagram in the effective theory generated by the operator in 
Eq.~\eqref{eq:scontact}
gives zero
on account of the vanishing of 
$c_{-2}$. 
Thus the finite volume correction for a point-like scalar remains universal in the effective theory.

There are operators of lower mass dimension in the contact Lagrangian, 
such as 
\begin{equation}
\cL_c (\phi, \chi)
=
- 
\mathcal{C}
\frac{Q^2 }{4 M^2}
\left(
\phi^\dagger 
\chi
\right)
\left(
\chi^\dagger  
\phi
\right)
\label{eq:scontact2}
,\end{equation}
that produce finite volume contributions to the electromagnetic mass. 
The contact operator above leads to an electromagnetic mass shift of the scalar
\begin{equation}
\frac{\Delta M^c_\Phi}{M}
=
\frac{\mathcal{C} \, Q^2}{8 (M L)^3}
\, c_0
,\end{equation}
which is non-vanishing due to the zero-mode subtraction. 
The coefficient 
$\mathcal{C}$
of this lower-dimensional contact operator, 
however, 
requires matching scalar QED at one-loop, 
for which
$\mathcal{C} \sim \alpha^2$, 
and the corresponding finite volume correction from antiparticles scales as 
$\cO(\alpha^2 L^{-3})$, 
which is beyond our present considerations.

In the spin-$\frac{1}{2}$ case, 
the relevant particle-antiparticle contact operator was identified in%
~\cite{Fodor:2015pna}, 
and we summarize their findings. 
For a non-relativistic spinor field 
$\psi$
and its antiparticle 
$\chi$, 
the leading contact operator has the form 
\begin{equation}
\cL^{(s)}_c (\psi, \chi)
= 
-
\frac{Q^2 e^2}{4 M^2}
(\psi^\dagger \bm{\sigma} \chi) 
\cdot
(\chi^\dagger \bm{\sigma} \psi)
,\end{equation}
where the coefficient has been determined by matching to QED at tree level%
~\cite{Bodwin:1994jh}.
The antiparticle tadpole generated from this spin-dependent contact interaction reproduces the missing contribution to the electromagnetic mass, 
Eq.~\eqref{eq:missingspin}. 
Having accounted for antiparticle contributions to the finite volume electromagnetic mass of point-like particles, 
we now extend our consideration to composite particles using NRQED.

\section{Composite Hadrons}
\label{sec:comp}

Given our demonstration for point-like particles, 
it is natural to extend the discussion to composite hadrons. 
In the context of non-relativistic effective field theories, 
we begin with composite scalars, 
and then move on to composite spinors. 
In each case, 
we determine the finite volume electromagnetic mass shift to 
$\cO(\alpha L^{-5})$, 
which is one order beyond that computed in%
~\cite{Davoudi:2014qua}, 
and include antiparticle contributions, 
which are shown to depend on squares of time-like electromagnetic form factors evaluated at threshold.

\subsection{Composite Scalars}

\begin{table}
\caption{Low-energy constants for composite hadrons in sNRQED and NRQED.
Listed are those coefficients required in the present study.}
\begin{center}
\begin{tabular}{|c||c|c|}
\hline
Constant 
&
sNRQED
&
NRQED
\\
\hline
$c_F$
&
-
&
$Q + \kappa$
\\
\hline
$c_D$
&
$\frac{4}{3} M^2 < r^2>$
&
$Q+ \frac{4}{3} M^2 < r_E^2>$
\\
\hline
$c_M$
&
$\frac{2}{3} M^2 < r^2>$
&
$\frac{2}{3} M^2 < r^2_E> - \frac{1}{2} \kappa$
\\
\hline
$c_S$
&
-
&
$Q + 2 \kappa$
\\
\hline
$c_{A_1}$
&
$16 \pi M^3 \beta_M$
&
$Q^2 + 16 \pi M^3 \beta_M$
\\
\hline
$c_{A_2}$
&
$\frac{8}{3} Q M^2 < r^2> \phantom{sp}$
&
$\frac{8}{3} Q M^2 < r_E^2> + 2 \kappa^2$
\\
&
$\phantom{sp}- 32 \pi M^3 ( \alpha_E + \beta_M)$
&
$\phantom{sp}- 32 \pi M^3 ( \alpha_E + \beta_M)$
\\
\hline
\hline
\end{tabular}
\end{center}
\label{t:constants}
\end{table}

Scalar NRQED (sNRQED) consists of three parts:
the Lagrangian density for the particle field, 
that of the antiparticle field, 
and finally their contact interactions. 
Thus we write the sNRQED Lagrangian density in the form
\begin{eqnarray}
\cL_{\text{sNRQED}}
=
\cL (\phi, e)
+ 
\cL (\chi, -e)
+
\cL_c (\phi, \chi)
\label{eq:sNRQED}
,\end{eqnarray}
where the particle and antiparticle fields are described by the Lagrangian density
\begin{eqnarray}
\cL(\phi, e)
&=&
\phi^\dagger
\Bigg[
i \overset{\leftrightarrow}{D}_0 
+
\frac{\overset{\leftrightarrow}{\bm{D}} {}^2}{2M}
+ 
e c_D 
\frac{[\bm{\nabla} \cdot \bm{E} ]}{8M^2}
+ 
\frac{\overset{\leftrightarrow}{\bm{D}} {}^4}{8 M^3}
\notag \\
&&
+
i e c_M
\frac{\left\{ D^i, [\bm{\nabla} \times \bm{B} ]^i \right\}}{8 M^3}
+
e^2 c_{A_1} 
\frac{\bm{B}^2 - \bm{E}^2}{8M^3}
\notag \\
&&-
e^2 c_{A_2}
\frac{\bm{E}^2}{16 M^3}
+
e c_{X_1}
\frac{\left[ \bm{D}^2, \bm{D} \cdot \bm{E} + \bm{E} \cdot \bm{D} \right]}{16 M^4}
\notag \\
&&+ 
e c_{X_2} 
\frac{
\left\{ \bm{D}^2, [ \bm{\nabla} \cdot \bm{E} ] \right\}
}{16 M^4}
+ 
e c_{X_3}
\frac{[ \bm{\nabla}^2 \bm{\nabla} \cdot \bm{E}] }{16 M^4}
\notag \\
&&+
i e^2 c_{X_4}
\frac{ \left\{ \bm{D}^i, (\bm{E} \times \bm{B} )^i\right\} }{16M^4}
\Bigg] \phi
\label{eq:LNRQED}
.\end{eqnarray}
The constants appearing above have been determined in%
~\cite{Lee:2013lxa},  
and the ones relevant for the present work are given in 
Table~\ref{t:constants}.

To the order we work, 
only the leading contact interaction Lagrangian density is required.
To leading order in $\alpha$, 
the contact operator is given in Eq.~\eqref{eq:scontact}, 
however, 
the coefficient of this operator must be modified to reflect the composite nature of the hadron. 
Due to the perturbative nature of QED, 
it suffices to consider the one-photon annihilation diagram in 
Fig.~\ref{f:contact}
to perform the matching. 
Unlike the point-like case, 
the annihilation vertex is replaced by the full hadronic amplitude, 
for example
\begin{equation}
\langle \Phi(p) \Phi^\dagger(p') | J_\mu | 0 \rangle
= 
e (p' - p)_\mu
F_\Phi (s)
\label{eq:timelikeFFs}
,\end{equation}
where 
$s = (p'+p)^2 \geq 4 M^2$
is the photon's virtuality, 
and 
$F_\Phi(s)$
is the time-like electromagnetic form factor of 
$\Phi$.
Carrying out the computation of the annihilation diagram for a composite scalar and performing the threshold expansion leads us to the contact Lagrangian density
\begin{eqnarray}
\cL_c (\phi, \chi)
=
- \frac{e^2 |\mathcal{F}|^2}{16 M^4}
\left(
\phi^\dagger 
i\overset{\leftrightarrow}{\bm{D}}  \,
\chi
\right)
\cdot
\left(
\chi^\dagger  
i \overset{\leftrightarrow}{\bm{D}} \,
\phi
\right)
\label{eq:seftcontact}
,\end{eqnarray}
where 
$\mathcal{F} \equiv F_\Phi(s = 4 M^2)$
is the form factor at threshold. 
For a point-like particle, 
we accordingly recover the contact interaction in 
Eq.~\eqref{eq:scontact} 
from that in
Eq.~\eqref{eq:seftcontact}.

%
%
\begin{figure}
\epsfig{file=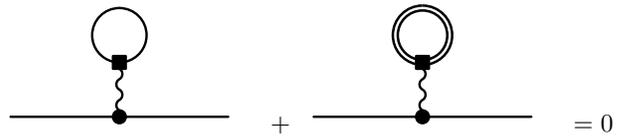,width=0.45\textwidth}
\caption{
Photon-mediated tadpole diagrams in the effective theory. 
The circles and squares denote the various photon couplings present in the NRQED Lagrangian density. 
For each possible square vertex, 
the sum of particle and antiparticle loops vanishes for a given circle vertex due to charge conjugation.  
}
\label{f:photad}
\end{figure}
%
%

With the specification of 
Eq.~\eqref{eq:sNRQED}, 
we have all ingredients necessary to compute the finite volume electromagnetic mass shift of the composite scalar particle to 
$\cO(\alpha L^{-5})$.
Repeating the calculation at 
$\cO(\alpha L^{-4})$, 
we recover the result derived in%
~\cite{Davoudi:2014qua}. 
We must augment this effective field theory calculation by considering two additional classes of diagrams.  
The first class consists of photon-mediated tadpoles, 
while the second class consists of antiparticle tadpoles. 
The photon-mediated tadpoles arise from the particle exchanging a photon with a particle or antiparticle loop, 
see Fig.~\ref{f:photad}. 
Both of these contributions are potentially non-vanishing in the effective theory, 
however, 
the sum of the two necessarily vanishes by charge conjugation invariance. 
Two photons coupling to particle and antiparticle loops are charge conjugation even and do not cancel;
however, 
these effects are 
$\cO(\alpha^2)$
and beyond our present considerations.

The antiparticle tadpole contribution has been determined above for a point-like particle, 
and it is trivial to modify this result for a composite scalar: 
$Q^2 \to |\mathcal{F}|^2$. 
This contribution is at 
$\cO(\alpha L^{-5})$, 
and while there are other such terms at this order, 
they cannot affect the finite volume mass because the accompanying shape coefficient vanishes, 
$c_{-2} = 0$. 
Thus there is no modification to the result of%
~\cite{Davoudi:2014qua}
one order beyond what was considered in that work. 
For completeness, 
we cite the 
$\cO(\alpha L^{-5})$ 
result here
\begin{eqnarray}
\Delta M_\phi
&=&
\Delta M^{(1,2)}
+
\frac{2 \pi  \alpha }{3 L^3}
Q < r^2>
-
\frac{4 \pi^2}{L^4}
\left(
\alpha_E 
+ 
\beta_M
\right)
c_{-1}
\notag \\
&&
+
\frac{8 \pi^3 \alpha}{3 M L^4}
Q < r^2>
c_{-1}
+ 
\cO(\alpha L^{-6})
.\end{eqnarray}

Notice that in the above expression, 
there is no contribution from anti-scalar modes. 
This result is actually true to all orders in 
$\cO(L^{-1})$, 
and to leading order in 
$\alpha$. 
As we have seen, 
such contributions arise from antiparticle tadpole diagrams generated by the contact interactions in the effective theory. 
The coefficients of these operators are determined from matching, 
which requires expanding the time-like electromagnetic form factor about threshold, 
see Eq.~\eqref{eq:timelikeFFs}.
Relative corrections in the threshold expansion of the annihilation amplitude always contain even powers of the momentum.
As a result, 
the contact operators in the effective theory for a composite scalar contain an even number of derivatives
(which also follows on account of invariance under parity). 
Based on power counting, 
the corresponding antiparticle tadpole diagrams are thus accompanied by the shape coefficients 
$c_{-2j}$, 
for 
$j = 1$, $2$, $\cdots$. 
Because all of these shape coefficients vanish, 
see Appendix~\ref{A:B}, 
there are no contributions from anti-scalar modes in the finite volume electromagnetic mass of a composite scalar to 
$\cO(\alpha)$. 
The same conclusion does not apply to 
$\cO(\alpha^2)$
finite volume corrections.

\subsection{Composite Spinors}

Now we consider finite volume corrections to the electromagnetic mass of a composite spin-half particle. 
By retaining the leading contribution from composite spin-half antiparticles in NQRED, 
we derive finite volume corrections valid to 
$\cO(\alpha L^{-5})$. 
This requires matching four-fermion operators for a composite spinor and generalizes the computation of%
~\cite{Fodor:2015pna}.

A composite spinor hadron can be described at low-energies by the NRQED 
Lagrangian density. 
There are three terms necessary: 
the Lagrangian density for the spin-half particle
$\psi$, 
that for the spin-half antiparticle
$\chi$, 
and the Lagrangian density containing their contact interactions. 
Writing down these three terms, 
we have
\begin{equation}
\cL_{\text{NRQED}}
=
\cL^{(s)} ( \psi, e) 
+ 
\cL^{(s)} (\chi, -e)
+ 
\cL^{(s)}_c ( \psi, \chi)
\label{eq:NRQED}
,\end{equation}
where the superscript is used to denote that the Lagrangian density is that relevant for a spinor field. 
For the spin-half particle field, 
the Lagrangian density 
$\cL^{(s)}$ 
can be decomposed into spin-independent and spin-dependent terms
\begin{equation}
\cL^{(s)}(\psi, e)
=
\cL(\psi, e)
+ 
\cL^{(\bm{\sigma})}(\psi,e)
,\end{equation}
where the spin-independent part 
$\cL(\psi, e)$
has precisely the form given in 
Eq.~\eqref{eq:LNRQED}.

The spin-dependent NRQED Lagrangian density reads
\begin{eqnarray}
\cL^{(\bm{\sigma})}
(\psi, e)
&=&
e \, \psi^\dagger 
\Bigg[
c_F 
\frac{\bm{\sigma} \cdot \bm{B}}{2 M}
+ 
i c_S
\frac{\bm{\sigma} \cdot \left( \bm{D} \times \bm{E} - \bm{E} \times \bm{D} \right)}{8 M^2}
\notag \\
&&\phantom{sp}+
c_{W_1}
\frac{ \{ \bm{D}^2, \bm{\sigma} \cdot \bm{B} \}}{8 M^3} 
- 
c_{W_2}  
\frac{D^i \bm{\sigma} \cdot \bm{B} D^i}{4 M^3}
\notag \\
&&\phantom{sp}+
c_{p'p}
\frac{\{ \bm{\sigma} \cdot \bm{D}, \bm{B} \cdot \bm{D} \} }{8 M^3}
\Bigg] \psi
.
\end{eqnarray}
The coefficients of spin-dependent operators that are pertinent for our computation are given in 
Table~\ref{t:constants}.
The Lagrangian density for the spin-half antiparticle is related by charge conjugation to that of the spin-half particle, 
$\cL^{(s)} (\chi, -e) = \cL(\chi, -e) + \cL^{(\bm{\sigma})} (\chi, -e)$.

Unlike the contact interaction in the scalar case, 
the four-fermion contact interaction required in the spinor calculation appears at 
$\cO(M^{-2})$,
and hence makes a contribution to the electromagnetic mass in finite volume of 
$\cO(L^{-3})$. 
The leading contact operator generated at 
$\cO(\alpha)$
is contained in the Lagrangian density
\begin{equation}
\cL^{(s)} (\psi, \chi)
= 
-
\mathcal{C}^{(s)}
\frac{e^2}{4 M^2}
(\psi^\dagger \bm{\sigma} \chi) 
\cdot
(\chi^\dagger \bm{\sigma} \psi)
.\end{equation}
The coefficient of this operator, 
$\mathcal{C}^{(s)}$,
must be determined by matching the hadronic annihilation amplitude with that computed in NRQED. 
As with the scalar case, 
the annihilation process can be computed perturbatively in QED using the annihilation diagram shown in 
Fig.~\ref{f:contact}. 
The hadron one-photon annihilation vertex must be described non-perturbatively by time-like electromagnetic form factors, 
for example
\begin{equation}
< \psi(p) \overline{\psi} (p') | J_\mu | 0 \rangle
=
e
\overline{u} (p) 
\left[
F_1(s) \gamma_\mu 
+
\frac{i \sigma_{\mu \nu}q^\nu }{2M} 
F_2 (s)
\right]
v(p')
,\end{equation}
with 
$q_\mu = p'_\mu + p_\mu$, 
and 
$s = q^2 \geq 4M^2$. 
The matching computation is an extension of that in%
~\cite{Bodwin:1994jh}, 
and results in the value of the coefficient
$\mathcal{C}^{(s)} = | \mathcal{G}_M |^2$, 
where 
\begin{equation}
\mathcal{G}_M \equiv F_1 ( s = 4 M^2) + F_2 (s = 4 M^2)
,\end{equation}
is the time-like Sachs magnetic form factor evaluated at threshold. 
In the case of a point-like fermion, 
$\mathcal{G}_M \to Q$, 
and this ensures the result of%
~\cite{Fodor:2015pna}
is recovered.

Having determined the coefficient of the fermion-antifermion contact operator, 
we have all of the necessary terms of the NRQED Lagrangian density, 
Eq.~\eqref{eq:NRQED}, 
to perform the finite volume mass calculation to 
$\cO(\alpha L^{-5})$. 
The result is identical to that derived in%
~\cite{Davoudi:2014qua}
aside from the additional term at 
$\cO(\alpha L^{-3})$
arising from the antifermion tadpole shown in Fig.~\ref{f:anntad}.
As with the scalar case, 
the sum of photon-mediated tadpole diagrams shown in 
Fig.~\ref{f:photad} 
vanishes, 
as do such diagrams with a kinetic energy insertion. 
As a result, 
there are no 
$\cO(\alpha L^{-4})$
corrections due to antiparticle modes. 
One order beyond that computed in%
~\cite{Davoudi:2014qua}, 
we observe that all
$\cO(\alpha L^{-5})$
contributions are proportional to the shape coefficient 
$c_{-2}$ 
which vanishes. 
For completeness, 
we give the full result to 
$\cO(\alpha L^{-5})$
accuracy
\begin{widetext}
\begin{eqnarray}
\Delta M_\psi
&=&
\Delta M^{(1,2)}
+
\frac{\pi \alpha}{M^2 L^3}
\Big[
\frac{2}{3} M^2 Q <r_E^2>
+
\frac{1}{2} Q^2
+
\frac{3}{2}
| \mathcal{G}_M|^2 
+
(Q+ \kappa)^2
\Big] 
- 
\frac{4 \pi^2}{L^4}
\left( \alpha_E + \beta_M \right)
c_{-1}
\notag\\
&&
+
\frac{\pi^2 \alpha}{M^3 L^4}
\left[
Q
\left(
\frac{4}{3} M^2 < r_E^2>
-
\kappa
\right)
- 
\kappa (Q + \kappa)
\right]
c_{-1}
+ 
\cO(\alpha L^{-6})
\label{eq:answer}
.\end{eqnarray}
\end{widetext}
Notice that the finite-volume effect from antiparticles in the case of spinor particles is due to infrared regularization, 
namely the subtraction of photon zero modes is accompanied by the need to remove loop contributions in the effective theory with vanishing antiparticle momentum%
~\cite{Fodor:2015pna}.
At higher orders in 
$L^{-1}$, 
however, 
there are no finite volume contributions from antiparticle modes at leading order in the fine-structure constant. 
As in the case of a composite scalar, 
this happens on account of parity invariance, 
which restricts the contact operators to have even mass dimension. 
The corresponding antiparticle tadpoles generated are hence accompanied by the shape coefficients 
$c_{-2j}$, 
with 
$j = 0$, $1$, $\cdots$. 
The only non-vanishing antiparticle tadpole on a spatial torus is that with 
$j=0$. 
This contribution has been accounted for in Eq.~\eqref{eq:answer}.

While the sole 
$\cO(\alpha)$
contribution from antiparticle modes is suppressed by two orders relative to the leading finite volume correction for charged spinors, 
the situation is different for neutral spinors. 
The leading finite-volume correction for a neutral spinor receives contributions from the particle-antiparticle contact interaction. 
In the case of the neutron, 
for example, 
we have 
\begin{eqnarray}
\Delta M_n
&=&
\frac{\pi \alpha}{M^2 L^3}
\left( 
\kappa_n^2 
+
\frac{3}{2} | \mathcal{G}^n_M|^2
\right)
- 
\frac{4 \pi^2}{L^4}
\left( \alpha_E^n + \beta_M^n \right)
c_{-1}
,\notag \\
\label{eq:neutron}
\end{eqnarray}
which is valid to 
$\cO(\alpha L^{-5})$. 
As the time-like form factors of the neutron are not very well known experimentally%
~\cite{Denig:2012by}, 
careful study of the finite volume effects for the neutron electromagnetic mass could, 
in principle, 
constrain 
the magnetic form factor at threshold.
In practice, 
determining a small finite-volume effect through the electromagnetic mass of the neutron is quite challenging, 
as the determination of baryon mass splittings already has considerable computational demands%
~\cite{Borsanyi:2014jba}.

%

\section{Related Chiral Concerns}
\label{sec:chiral}

It is instructive to contrast the electromagnetic mass of the nucleon with the QCD self-energy of the nucleon. 
We verify that finite volume contributions from the strong nucleon-antinucleon annihilation process vanish due 
to the locality of finite volume QCD.

\begin{figure}
\epsfig{file=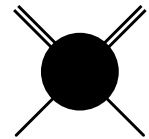,width=0.10\textwidth}
\caption{
Strong nucleon-antinucleon annihilation.
}
\label{f:nuked}
\end{figure}

At low energies, 
one can use chiral perturbation theory to describe the long-range physics of the nucleon. 
As a result, 
this effective field theory gives predictions for the volume dependence of the nucleon mass in terms of the 
pion Compton wavelength, 
$m_\pi^{-1}$. 
For most lattice QCD applications, 
one works in the $p$-regime, 
where
$m_\pi L \gg 1$,
which ensures that zero modes of the pion field do not become strongly coupled%
~\cite{Gasser:1987zq}. 
Strong coupling of pion zero modes is quite similar to zero modes in finite volume QED, 
which can be exemplified by giving the photon a small mass%
~\cite{Endres:2015gda}.
The natural question to address in light of our work is whether there are annihilation contributions missing from standard chiral perturbation theory analyses of the volume dependence of the nucleon mass%
~\cite{AliKhan:2003cu,Beane:2004tw}. 
Nucleon-antinucleon annihilation is described by the amplitude depicted in 
Fig.~\ref{f:nuked}. 
Unlike the electromagnetic annihilation process, 
the strong interaction annihilation is not dominated by one-pion exchange. 
The nucleon-pion interactions are weak only at low energy, 
whereas the electromagnetic interaction remains weak well above the pair-production threshold, 
$q^2 \geq 4 M^2$. 
The strong annihilation process can, 
however, 
be described by local operators in chiral perturbation theory. 
Momentum-independent operators, 
such as the spin/isospin independent contact operator
\begin{equation}
\cO
= 
\frac{1}{4 M^2}
\left( \psi^\dagger \chi \right)
\left( \chi^\dagger \psi \right)
,\end{equation}
would appear to give the leading contribution from the strong annihilation process. 
In infinite volume, 
such contact operators lead to an infinite renormalization of the nucleon mass, 
which can be absorbed by local nucleon operators in chiral perturbation theory. 
At finite volume, 
there is the possibility of a finite counter-term  from the antinucleon tadpole diagram.
Due to momentum independence, 
however, 
the contributions from all such operators are proportional to
\begin{equation}
\left(
\frac{1}{L^3} \sum_{\bm{n}}
- 
\int \frac{d\bm{q}}{(2 \pi)^3}
\right)
1
=
0.
\end{equation}
The vanishing of this regulated sum depends on the retention of antinucleon zero modes. 
In the 
$\epsilon$-regime, 
where
$m_\pi L \ll 1$, 
a separate analysis is required.

%
%
%
%

Antinucleon loops could contribute to the finite volume shift of the nucleon mass through higher-dimensional contact interactions. 
Operators of the form
\begin{equation}
\mathcal{O}
=
\frac{1}{16 M^4}
\left(
\psi^\dagger 
i\overset{\leftrightarrow}{\bm{D}}  \,
\chi
\right)
\cdot
\left(
\chi^\dagger  
i \overset{\leftrightarrow}{\bm{D}} \,
\psi
\right)
,\end{equation}
for example, 
would lead to finite volume contributions 
$\propto
M^{-4} L^{-5}
c_{-2}
$,
where the constant of proportionality depends the annihilation amplitudes at threshold in the possible spin/isospin channels. 
Such higher-dimensional operators with an even number of derivatives, 
however,
lead to mode sums involving the shape coefficients
$c_{- 2 j}$, 
which vanish. 
Thus provided one remains in the 
$p$-regime of chiral perturbation theory, 
there are no finite volume effects from any antinucleon modes. 
This is to be expected from the locality of finite volume QCD.

\section{Summary}
\label{sec:summary}

Above we consider the effect of antiparticle contributions in low-energy effective field theories. 
Interactions of non-relativistic particles and antiparticles can be described in NRQED, 
such as those required in the treatment of positronium.
In applying non-relativisitic effective theories to single hadrons,
however, 
it would seem that such short-distance antiparticle contributions only show up at low energies in the form of 
counter-terms.  
In infinite volume, 
this is precisely the case as can be demonstrated from computing annihilation tadpole diagrams. 
Such diagrams, 
however, 
can lead to non-vanishing effects at finite volume. 
In the non-local formulation of finite volume QED, 
antiparticles make finite contributions to the electromagnetic mass of hadrons with spin at 
$\cO(\alpha)$.

To make the discussion of these points concrete, 
we consider the case of point-like particles in QED. 
Antiparticle modes are exhibited in the covariant computation using the method of regions. 
In order for the power-law finite volume effect in the non-relativistic description of point-like particles to agree with the corresponding QED results, 
we require the addition of particle-antiparticle contact interactions. 
The strength of such interactions can be determined from matching conditions, 
and power counting arguments naturally explain why no antiparticle contributions exist in the scalar case, 
whereas such contributions are required in the spinor case.

We extend the considerations of%
~\cite{Fodor:2015pna}
to the case of composite particles.
In the case of a composite scalar, 
the matching condition is generalized,
however, 
symmetry and power counting arguments allow one to conclude that anti-scalar modes do not contribute to the finite volume effect at leading order in $\alpha$. 
Due to the vanishing of the shape coefficient
$c_{-2}$, 
we additionally find that results derived for composite scalars to 
$\cO(\alpha L^{-4})$ in%
~\cite{Davoudi:2014qua}
are valid to 
$\cO(\alpha L^{-5})$. 
For composite spinors, 
on the other hand,
antiparticle contributions modify results at 
$\cO(\alpha L^{-3})$;
and, 
arise from the zero-mode subtraction utilized to regulate QED in the infrared. 
Unlike the point-like spinor case, 
however,
the contribution from antiparticle modes is encoded in matching the full electromagnetic annihilation amplitude. 
This requires non-perturbative QCD physics, 
and we find that the finite volume effect is sensitive to the square of the 
Sachs magnetic form factor at threshold. 
From precision study of the finite volume dependence of the neutron electromagnetic mass, 
for example,
there is the amusing possibility to bound the time-like magnetic form factor at threshold because such 
$\cO(\alpha L^{-3})$ effects
are leading order for neutral particles, 
see Eq.~\eqref{eq:neutron}.
In the course of our investigation, 
we found that computations for composite spinors to 
$\cO(\alpha L^{-4})$
are valid to one order higher due to the identical vanishing of the shape coefficient
$c_{-2}$
appearing in
$\cO(\alpha L^{-5})$
finite volume effects.

Finally we assess the contribution of antiparticle modes in finite volume calculations of the nucleon mass. 
In chiral perturbation theory, 
the dominant finite volume effects arise from long-range pions that wind around the lattice volume. 
Nucleons winding around the lattice volume are exponentially suppressed relative to these. 
Antinucleon contributions at finite volume are shown to be absent due to the retention of pion zero modes, 
and hence the locality of finite volume QCD.

\begin{acknowledgments}
We gratefully acknowledge support for this work from the U.S.~National Science Foundation,
under Grant No.~PHY$15$-$15738$.
The work of BCT is additionally supported by a joint 
The City College of New York--RIKEN/Brookhaven Research Center fellowship,
and a grant from the Professional Staff Congress of The CUNY. 
BCT thanks M.~J.~Savage for discussions concerning this topic.
We also thank 
A.~Portelli and A.~Ramos for pointing out a crucial missing ingredient in an earlier version of this note. 
\end{acknowledgments}

\appendix

\section{Power-Law and Exponential Volume Effects}
\label{A:A}

In this Appendix, 
we evaluate the one-loop finite volume corrections to the point-like scalar and spinor electromagnetic masses. 
To accomplish this, 
we follow%
~\cite{Borsanyi:2014jba},
and separate the power-law from exponential behavior in 
$M L$. 
Only the leading exponential behavior is quoted in the main text.

In the case of point-like scalars and spinors, 
we write both finite volume corrections in the form
\begin{eqnarray}
\Delta M
= 
\Delta M^{(1,2)}
+ 
\delta (\Delta M)
.\end{eqnarray}
For point-like scalars, 
the correction beyond the universal one, 
which is defined as
$\delta (\Delta M_\Phi)$,
vanishes exponentially fast in the volume. 
From 
Eq.~\eqref{eq:demo}, 
this correction is given by
\begin{equation}
\delta (\Delta M_\Phi)
=
\frac{i Q^2 e^2}{2 M}
\int \frac{d q_0}{2 \pi}
\sumint
\frac{2 M + q_0}{(q_0 + i \epsilon) (2 M q_0 + q^2 + i \epsilon)}
,\end{equation}
and accordingly does not contain singularities associated with propagation of the photon. 
Furthermore, 
the zero-mode contribution to 
$\delta (\Delta M_\Phi)$
vanishes, 
and the sum over non-zero momentum modes can be trivially extended to all modes. 
Carrying out the integration over 
$q_0$
and utilizing the Poisson summation formula to perform a Fourier transform, 
we arrive at
\begin{equation}
\delta(\Delta M_\Phi)
=
-
\frac{Q^2 \alpha}{2 \pi M} 
\int_M^\infty
d \Lambda
(2 M - \Lambda)
\sum_{\vec{\nu} \neq \vec{0}}
\texttt{K}_0
( \Lambda L | \vec{\nu} | )
,\end{equation}
where
$\texttt{K}_0(x)$
is a modified Bessel function of the second kind. 
Here, 
the absence of 
$\vec{\nu} = \vec{0}$
from the winding number summation 
ensures that there are no contributions which survive the infinite volume limit. 
The asymptotic behavior of the Bessel function, 
$\texttt{K}_0(x) \sim \sqrt{\frac{\pi}{2x}} e^{ - x}$
for 
$x \gg 1$, 
leads to the leading exponential correction given in 
Eq.~\eqref{eq:scalarL},
which includes the periodic images with 
$| \vec{\nu} \, | = 1$.

For point-like spinors, 
on the other hand, 
the finite volume correction beyond the universal one, 
$\delta (\Delta M_\Psi)$, 
contains both power-law and exponential dependence on 
$L$. 
The residual power-law dependence is due to non-locality introduced by the photon zero-mode subtraction.
From Eq.~\eqref{eq:demo2}, 
we find the expression
\begin{equation}
\delta (\Delta M_\Psi)
=
\frac{i Q^2 e^2}{M}
\int \frac{d q_0}{2 \pi}
\sumint
\frac{M - q_0}{(q_0 + i \epsilon) (2 M q_0 + q^2 + i \epsilon)}
,\end{equation}
which has a form quite similar to the scalar case. 
Analogous to that case,
we perform the 
$q_0$
integration; 
however, 
in order to utilize the Poisson summation formula, 
we must add and subtract the non-vanishing zero mode contribution. 
This is the source of power-law volume dependence in 
$\delta (\Delta M_\Psi)$, 
and the full result takes the form
\begin{eqnarray}
\delta (\Delta M_\Psi)
&=&
\frac{3 \pi Q^2 \alpha}{M^2 L^3}
-
\frac{Q^2 \alpha}{\pi M} 
\int_M^\infty
d \Lambda
(M + \Lambda)
\sum_{\vec{\nu} \neq \vec{0}}
\texttt{K}_0
( \Lambda L | \vec{\nu} | )
.\notag \\
\end{eqnarray}
The asymptotic expansion of the modified Bessel function yields the leading exponential correction to the point-like spinor mass presented in 
Eq.~\eqref{eq:pointspin}.

\section{Shape Coefficients with Zero-Mode Subtraction}
\label{A:B}

In this Appendix, 
we evaluate the shape coefficients, 
$c_j$, 
which are formally defined by
\begin{equation}
c_j
=
\sum_{\bm{n} \neq \bm{0}}
\frac{1}{\, |\bm{n}|^j}
- 
\int \frac{d \bm{n}}{\, |\bm{n}|^j}
\notag 
.\end{equation}
These mode number sums arise in finite volume computations of QED with photon zero-mode subtraction, 
and have been employed in the main text. 
The sum and integral are each divergent in the ultraviolet, 
however, 
their difference is finite in a large class of ultraviolet regularization schemes.  
For this reason, 
we can treat the ultraviolet regularization as implicit. 
To evaluate these coefficients, 
we add a small mass term to regulate the behavior in the infrared, 
and work in $d$ spatial dimensions.
Thus we consider
\begin{equation}
c_j^{(d)}
=
\lim_{\mu \to 0}
\left[
\sum_{\bm{n} \neq \bm{0}}
\frac{1}{ [ \bm{n}^2 + \mu^2  ]^{j/2}}
- 
\int \frac{d^d \bm{n}}{[ \bm{n}^2 + \mu^2  ]^{j/2}}
\right]
,\end{equation}
with 
$c_j = c_j^{(3)}$
the desired quantities of interest. 
Following%
~\cite{Hasenfratz:1989pk},%
\footnote{
The same conclusions are reached using  
L\"uscher's zeta-function, 
as rigorously detailed in%
~\cite{Luscher:1986pf}. 
} 
we consider the coefficients as functions of  
$j>0$, 
and employ analytic continuation for 
$j \leq 0$.

We write these coefficients in terms of two contributions 
$c_j^{(d)} = a_j^{(d)} + b_j^{(d)}$. 
The 
$a_j^{(d)}$
are all non-singular in the infrared 
\begin{eqnarray}
a_j^{(d)}
&=&
\int_0^1 ds
\frac{s^{ - \frac{j}{2} - 1} + s^{\frac{j - d}{2}-1} }{\pi^{- \frac{j }{2}} \Gamma( \frac{j}{2}) }
\left[
\vartheta_3 
\left( 0, e^{ - \frac{\pi}{s}} \right)^d
-
1
\right]
,\quad 
\end{eqnarray}
with 
$\vartheta_3(z,q)$
is the Jacobi elliptic-theta function,
and can readily be continued to  
$j \leq 0$
because the quantity in brackets vanishes at 
$s = 0$
faster than any power of $s$. 
The 
$b_j^{(d)}$
coefficients, 
on the other hand,
potentially contain infrared singularities, 
and are given by
\begin{eqnarray}
b_j^{(d)}
&=&
\lim_{\mu \to 0}
\left[
\frac{\mathcal{B}_j (\mu) - \mathcal{B}_{j - d} (\mu)}{\pi^{-  \frac{j }{2}} \Gamma( \frac{j}{2}) }
-
\frac{1}{\mu^j}
\right]
,\end{eqnarray}
where the last term is the subtraction of the photon zero mode, 
and the incomplete gamma-function 
$\mathcal{B}_j(\mu)$
has the definition
\begin{eqnarray}
\mathcal{B}_j(\mu)
&=&
\int_1^\infty ds \,
s^{ \frac{j}{2} - 1}
e^{ - \pi s \mu^2}
\notag \\
&=&
\frac{
\Gamma \left( \frac{j}{2} \right)
}
{\left( \pi \mu^2 \right)^{\frac{j}{2}}
}
-
\sum_{n=0}^\infty
\frac{1}{n!}
\left( - \pi \mu^2 \right)^n
\frac{1}{n + \frac{j}{2}}
.
\end{eqnarray}
From explicit evaluation, 
we see the infrared singular behavior of the coefficients
\begin{equation}
b_{j \geq d}^{(d)}
=
\lim_{\mu \to 0}
\begin{cases}
\left(
\frac{1}{\mu}
\right)^{j - d}, 
& 
j > d
\\
\log \mu, 
& 
j = d
\end{cases}
.\end{equation}
For any charged particle, 
the largest possible value for 
$j$
entering the effective theory computation of the self energy is 
$j = 2$,
and arises at leading order in the non-relativistic expansion. 
Not surprisingly, 
we find that the photon zero-mode subtraction eliminates all infrared singularities provided that we work in 
$d > 2$
spatial dimensions.

For all values of 
$j$
such that 
$j < d$,
the coefficients 
$b_j^{(d)}$ 
are finite%
~\cite{Hasenfratz:1989pk,Luscher:1986pf}. 
Despite the many possible special cases depending on the sign of 
$j$, 
and the evenness or oddness of both
$j$
and
$d$, 
we find that the coefficients can be compactly written in the form
\begin{eqnarray}
b_{j < d}^{(d)}
&=&
-
\frac{ \pi^{ \frac{j}{2}}}{ \Gamma( \frac{j}{2}) }
\frac{2d}{j ( d - j)}
,\end{eqnarray}
with a limit implied for 
$j = 0$, 
$-2$, 
$-4$, 
$\cdots$. 
In particular, 
the special case 
$j = 0$
has the corresponding value
$b_0^{(d)} = - 1$
in all spatial dimensions 
$d$. 
Combining with the non-singular piece, 
$a_0^{(d)} = 0$, 
we have
\begin{equation}
c_0^{(d)}
= - 1
.\end{equation}
Finally both 
$a_j^{(d)}$
and 
$b_j^{(d)}$ 
contributions vanish at negative even integers. 
This leads us to
\begin{equation}
c_j^{(d)} 
= 0, 
\quad \text{for} \quad 
j = -2, -4, \cdots
\label{eq:zero}
,\end{equation}
in all spatial dimensions 
$d$. 
In three spatial dimensions, 
one has a special relation between shape coefficients,
$c_2^{(3)} = \pi c_1^{(3)}$. 
In the main text, 
we have made use of the additional values: 
$c_1^{(3)} = - 2.83729$, 
and
$c_{-1}^{(3)} = -0.266596$.

As many of the conclusions drawn in the main text depend on the values derived in Eq.~\eqref{eq:zero}, 
we show the vanishing of these coefficients alternatively by using an explicit Gaussian ultraviolet regulator.
Introducing the Gaussian regularization and without a mass term, 
we have
\begin{equation}
c_{- 2 j}^{(d)}
=
\lim_{\lambda \to 0}
\left(
\sum_{\bm{n}}
-
\int d^d \bm{n}
\right)
\bm{n}^{2j}
e^{ - \lambda \bm{n}^2}
.\end{equation}
Notice that the mode summation has been extended to include the zero mode, 
because the summand vanishes there.
Making use of the smoothness of the Gaussian 
we can write
\begin{eqnarray}
c_{- 2 j}^{(d)}
&=&
\lim_{\lambda \to 0}
\left( - \frac{\partial}{\partial \lambda} \right)^j
\left(
\sum_{\bm{n}}
-
\int d^d \bm{n}
\right)
e^{ - \lambda \bm{n}^2}
\notag \\
&=&
\sum_{\bm{\nu} \neq \bm{0}}
\lim_{\lambda \to 0}
\left( - \frac{\partial}{\partial \lambda} \right)^j
\left(\frac{\pi}{\lambda}\right)^{\frac{d}{2}}
e^{ - \pi^2 \bm{\nu}^2 / \lambda}
=0
,\quad \end{eqnarray}
where we have used 
$\lim_{\lambda \to 0} \lambda^\alpha e^{ - \beta / \lambda} = 0$, 
for any
$\alpha$
and for
$\beta > 0$. 
The value of 
$c_{0}^{(d)}$
can also be obtained from Gaussian regularization in the limit that 
$j \to 0$.
Notice the above expression produces the quantity
$c_{0}^{(d)} + 1 = 0$ 
because the zero mode has been included in the summation.

Separation of the ultraviolet behavior and volume dependence holds for a large class of regulators, 
such as dimensional regularization and smooth cutoffs.
Sharp ultraviolet cutoffs,  
by contrast,
tend to produce power divergences in finite volume corrections
which spoil the separation of ultraviolet from infrared physics%
~\cite{Hasenfratz:1989pk}.
In hard-cutoff regularization, 
for example, 
we have the conflicting result
\begin{equation}
\hat{c}_0^{(1)}
=
\lim_{\Lambda \to \infty}
\left[
-1
+
\sum_{n=-\Lambda}^\Lambda 
1
-
\int_{-\Lambda}^\Lambda
dn 
\, 1
\right]
=
0
\neq 
-1
,\end{equation}
where the first term is the explicit subtraction of the zero mode. 
The hard cutoff can be ruled out on physical grounds, 
however, 
as further shape coefficient exhibit ultraviolet singularities, 
e.g.,
\begin{equation}
\hat{c}_{-2}^{(1)}
=
\lim_{\Lambda \to \infty}
\left[
\sum_{n = -\Lambda}^{\Lambda}
n^2 
- 
\int_{-\Lambda}^\Lambda
dn \, n^2
\right]
=
\lim_{\Lambda \to \infty}
\Lambda^2
=
\infty
.\end{equation}
This behavior is physically unacceptable because the infinite volume limit of quantities can no longer be taken independent of the ultraviolet regularization.

\newpage

\bibliography{spinor}

\end{document}